\documentclass[aps,showpacs,preprint,superscriptaddress]{revtex4}
\usepackage{graphicx}
\usepackage{amsmath}
\usepackage{bbm}

\begin{document}

\title{Mass shift effects in nonperturbative multiphoton pair production for arbitrary polarized electric fields}

\author{Z. L. Li}
\author{D. Lu}
\affiliation{Key Laboratory of Beam Technology and Materials Modification of the Ministry of Education, College of Nuclear Science and Technology, Beijing Normal University, Beijing 100875, China}
\author{B. F. Shen}
\affiliation{Shanghai Institute of Optics and Fine Mechanics, Chinese Academy of Sciences, Shanghai 201800, China}
\author{L. B. Fu}
\author{J. Liu}
\affiliation{National Laboratory of Science and Technology on Computational Physics,
Institute of Applied Physics and Computational Mathematics, Beijing 100088, China}
\author{B. S. Xie\footnote{Corresponding author: bsxie@bnu.edu.cn}}
\affiliation{Key Laboratory of Beam Technology and Materials Modification of the Ministry of Education, College of Nuclear Science and Technology, Beijing Normal University, Beijing 100875, China}
\affiliation{Beijing Radiation Center, Beijing 100875, China}

\date{\today}

\begin{abstract}
The mass shift effects in multiphoton pair production of a nonperturbative nature for arbitrary polarized electric fields are investigated numerically by employing the real-time Dirac-Heisenberg-Wigner formalism, and theoretically by proposing an effective energy concept. It is found that the theoretical results are agreement with the numerical ones very well. It is the first time to consider the roles of the momenta of created particles and the polarizations of external fields played in the mass shift effects. These results can deepen the understanding of pair production in the nonperturbative threshold regime. Moreover, the distinct mass shift effects are observable in the forthcoming experiments and can be used as a probe to distinguish the electron-positron pair production from other background events.

\end{abstract}

\maketitle

\textit{Introduction.}---On the basis of one of the theoretical predictions of quantum electrodynamics, a vacuum in the presence of strong fields is unstable and will decay into electron-positron (EP) pairs \cite{Sauter,Heisenberg,Schwinger}. To experimentally observe this phenomenon, according to the calculation of Schwinger, the strength of external electric fields should be comparable to the very high critical electric field strength $E_{\mathrm{cr}}=m^2/e\sim1.32\times10^{16}\mathrm{V/cm}$, where $m$ is the electron rest mass and $e$ is the magnitude of electron charge (the units $\hbar=c=1$ are used). This electric field strength is far beyond what the current laboratories can achieve. However, some authors \cite{Brezin,Dunne} found that the EP pair production may be observed for a time-varying electric field with the electric field strength lower than the critical one. Furthermore, recent experiments are planning to achieve the laser fields $1$ or $2$ orders of magnitude lower than the critical electric strength in the high-intensity and ultrashort laser facilities such as the Extreme Light Infrastructure \cite{ELI} and the x-ray free electron laser (XFEL) \cite{Ringwald}. These theoretical and experimental developments again raise the hopes to realize an experimental detection of EP pair production from vacuum \cite{Piazza1}. To experimentally observe the Schwinger pair production, many catalytic mechanisms \cite{Bell,Piazza2,GVDunne,Bulanov,Monin} were put forward, such as the dynamically assisted Schwinger mechanism \cite{Schutzhold} and the multi-time-slit interference effects \cite{Akkermans}.

Comparing with the nonperturbative Schwinger pair production ($\gamma\ll1$), the perturbative multiphoton pair production ($\gamma\gg1$) in a laser field has been accomplished more than a decade ago at the Stanford Linear Accelerator Center via the collisions of a $46.6$ GeV electron beam with an intense optical laser pulse \cite{Burke}. Note that the two different process are divided by the well-known Keldysh adiabaticity parameter $\gamma=m\omega/(eE_0)$ \cite{Keldysh}, where $\omega$ and $E_0$ are the frequency and strength of external electric fields, respectively. Although these two different mechanisms have been well investigated, the intermediate regime ($\gamma\sim\mathcal{O}(1)$), i.e., the nonperturbative multiphoton pair production, is seldom considered \cite{Ruf,Mocken}, because there are no simple asymptotic formulae in this regime. However, as the nonperturbative multiphoton process contains not only the perturbative feature but also the nonperturbative nature, it becomes a very interesting research topic both in the theory and the experiment. Furthermore, there are many novel phenomena occurring in this regime, for instance, the effective mass signatures \cite{Kohlfurst2014}.

In fact the mass shift effects can be commonly seen when electrons pass through the plane wave fields \cite{Volkov}, the undulator fields \cite{McNeil}, and the general fields \cite{Dodin}. Its existence, universality, and detection in laser-particle scattering were studied in Ref. \cite{Harvey}. Recently, an effective mass model \cite{Kohlfurst2014} was put forward to interpret the mass shift effects in the EP pair yield varying with the laser frequency. However, there are still some unsolved issues: What are the roles of the momenta of created EP pairs played in the mass shift effects? By integrating over the full momentum space, what are the changes about the mass shift effects? Does the effective mass model still hold true?

In this paper, we focus our study on the mass shift effects in the nonperturbative multiphoton pair production for arbitrary polarized electric fields by numerically solving the real-time Dirac-Heisenberg-Wigner (DHW) formalism \cite{Bialynicki,Hebenstreit2010} as well as by theoretically proposing an effective energy. We will make the problems mentioned above clear and deepen the understanding of the mass shift effects in the multiphoton pair production of a nonperturbative nature. In addition, the effects of the polarizations of external electric fields on the mass shift effects is considered as well.

\textit{Arbitrary polarized fields.}---Under anticipated XFEL conditions, $E\lesssim 0.1E_{\mathrm{cr}}$, it is a good approximation to neglect the collision effect and the internal electric field since the EP pair yield and the back-reaction electric current are quite small. And because the spatial scales of the EP pair production are smaller than the spatial focusing scales of the laser pulse, the spatial effects are not significant. Therefore, we have the spatially homogeneous and time-dependent fields. For our studies, we focus on the EP pair production in a uniform and time-varying electric field of arbitrary polarization
\begin{equation}\label{eq1}
\mathbf{E}(t)=E_0\exp\Big(-\frac{t^2}{2\tau^2}\Big)\left[
                                             \begin{array}{c}
                                               \cos(\omega t+\phi) \\
                                               \delta\sin(\omega t+\phi) \\
                                               0 \\
                                             \end{array}
                                           \right],
\end{equation}
where $E_0$ is the maximal field strength, $\tau$ defines the pulse duration, $\omega$ is the laser frequency, $\phi$ is the carrier phase, and $-1\leq\delta\leq1$ represents the polarization. Note that the magnetic effects are ignored since we focus on the standing-wave field formed by two counter propagating laser pulses with appropriate polarization. For convenience, we set $\tau=100$ and $\phi=0$ throughout this paper.

\textit{Dirac-Heisenberg-Wigner formalism.}---Our following numerical results are based on the DHW formalism which has been used to study vacuum pair production in Refs. \cite{Bialynicki,Hebenstreit2010} for different electric fields. We start with the equal-time density operator of two Dirac field operators in the Heisenberg picture,
\begin{eqnarray}\label{eq2}
\hat{\mathcal{C}}_{\alpha\beta}(\mathbf{x},\mathbf{y},t)=&&e^{-ie\int^{1/2}_{-1/2}
\mathbf{A}(\mathbf{x}+\lambda \mathbf{y},t)\cdot \mathbf{y} d\lambda}\nonumber\\
&&\times\Big[\hat{\Psi}_\alpha\Big(\mathbf{x}+\frac{\mathbf{y}}{2},t\Big),
\hat{\bar{\Psi}}_\beta\Big(\mathbf{x}-\frac{\mathbf{y}}{2},t\Big)\Big],
\end{eqnarray}
with the center-of-mass coordinate $\mathbf{x}=(\mathbf{x}_1+\mathbf{x}_2)/2$ and the relative coordinate $\mathbf{y}=\mathbf{x}_1-\mathbf{x}_2$. Note that the factor before the commutator is a Wilson-line factor used to keep gauge invariance, and the integration path of the vector potential $\mathbf{A}$ is a straight line chosen to introduce a clearly defined kinetic momentum $\mathbf{p}$ . Moreover, we have employed a Hartree approximation for the electromagnetic field and chosen the temporal gauge $A_0=0$. The Wigner operator is defined as the Fourier transformation of Eq. (\ref{eq2}) with respect to the relative coordinate $\mathbf{y}$, and its vacuum expectation value gives the Wigner function
\begin{eqnarray}\label{eq3}
\mathcal{W}(\mathbf{x},\mathbf{p},t)=-\frac{1}{2}\int d^3ye^{-i\mathbf{p}\cdot\mathbf{y}}\langle0|\hat{\mathcal{C}}(\mathbf{x},\mathbf{y},t)|0\rangle.
\end{eqnarray}
Decomposing the Wigner function in terms of a complete basis set $\{\mathbbm{1},\gamma_5,\gamma^\mu,\gamma^\mu\gamma_5,\sigma^{\mu\nu}:=\frac{i}{2}
[\gamma^\mu,\gamma^\nu]\}$, we have
\begin{equation}\label{eq4}
\mathcal{W}(\mathbf{x},\mathbf{p},t)=\frac{1}{4}(\mathbbm{1}\mathbbm{s}
+i\gamma_5\mathbbm{p}+\gamma^\mu\mathbbm{v}_\mu
+\gamma^\mu\gamma_5\mathbbm{a}+\sigma^{\mu\nu}\mathbbm{t}_{\mu\nu}),
\end{equation}
with sixteen real Wigner components, scalar $\mathbbm{s}(\mathbf{x},\mathbf{p},t)$, pseudoscalar $\mathbbm{p}(\mathbf{x},\mathbf{p},t)$, vector $\mathbbm{v}(\mathbf{x},\mathbf{p},t)$, axialvector $\mathbbm{a}(\mathbf{x},\mathbf{p},t)$, and tensor $\mathbbm{t}(\mathbf{x},\mathbf{p},t)$. Inserting the decomposition into the equation of motion for the Wigner funtion, one can obtain a partial differential equation (PDE) system for the sixteen Wigner components \cite{Bialynicki}. Furthermore, for the spatially homogeneous and time-dependent electric fields mentioned above, by using the method of characteristics, or simply, replacing the kinetic momentum $\mathbf{p}$ by $\mathbf{q}-e\mathbf{A}(t)$ with the well-defined canonical momentum $\mathbf{q}$, the PDE system for the sixteen Wigner components can be reduced to an ordinary differential equation system for the ten nontrivial Wigner components $\mathbbm{w}(\mathbf{q},t)=(\mathbbm{s},\mathbbm{v},\mathbbm{a},\mathbbm{t}_1:
=2\mathbbm{t}^{i0}\mathbf{e}_i)^\textsf{T}(\mathbf{q},t)$,
\begin{equation}\label{eq5}
\dot{\mathbbm{w}}(\mathbf{q},t)=\mathcal{H}(\mathbf{q},t)\mathbbm{w}(\mathbf{q},t),
\end{equation}
where the dot denotes a total time derivative, $\mathcal{H}(\mathbf{q},t)$ is a $10\times10$ matrix.
The one-particle distribution function is defined as
\begin{equation}\label{eq6}
f(\mathbf{q},t)=\frac{1}{2}\mathbbm{e}^\textsf{T}_1\cdot[\mathbbm{w}(\mathbf{q},t)
-\mathbbm{w}_{\mathrm{vac}}(\mathbf{q},t)],
\end{equation}
where $\mathbbm{w}_{\mathrm{vac}}(\mathbf{q},t)=(\mathbbm{s}_{\mathrm{vac}}~\mathbbm{v}_{\mathrm{vac}}~ \mathbf{0} ~~\mathbf{0})^\textsf{T}$, $\mathbbm{s}_{\mathrm{vac}}=-2m/\Omega(\mathbf{p})|_{\mathbf{p}\rightarrow \mathbf{q}-e\mathbf{A}(t)}$, $\mathbbm{v}_{\mathrm{vac}}=-2\mathbf{p}/\Omega(\mathbf{p})|_{\mathbf{p}\rightarrow \mathbf{q}-e\mathbf{A}(t)}$, $\Omega(\mathbf{p})|_{\mathbf{p}\rightarrow \mathbf{q}-e\mathbf{A}(t)}=\sqrt{m^2+[\mathbf{q}-e\mathbf{A}(t)]^2}$ is the total energy of electrons, and $\mathbbm{e}_1=-1/2~\mathbbm{w}_{\mathrm{vac}}$ is one of the basis of the ten-component vector $\mathbbm{w}$. Notice that the vacuum solution is $\mathbbm{w}_{\mathrm{vac}}(\mathbf{q},t_{\mathrm{vac}})$.

In order to precisely obtain the distribution function $f$, we adopt the trick used in \cite{Blinne}. Decomposing the Wigner components as  $\mathbbm{w}=2(f-1)\mathbbm{e}_1+\mathcal{F}\mathbbm{w}_9$ with an auxiliary nine-component vector $\mathbbm{w}_9$ and a $10\times9$ matrix
$\mathcal{F}=\left(
               \begin{array}{cccccc}
                 -\mathbf{p}^\textsf{T}/m ~~ \mathbf{0} \\
                 ~~~\mathbbm{1}_9 \\
               \end{array}
             \right)\Big|_{\mathbf{p}\rightarrow \mathbf{q}-e\mathbf{A}(t)},
$ and applying Eq. (\ref{eq5}), we have
\begin{eqnarray}\label{eq7}
&\dot{f}=1/2~\dot{\mathbbm{e}}_1^\textsf{T} \mathcal{F}\mathbbm{w}_9,&  \nonumber\\
&\dot{\mathbbm{w}}_9=\mathcal{H}_9\mathbbm{w}_9+2(1-f)\mathcal{G}\dot{\mathbbm{e}}_1,&
\end{eqnarray}
where $\mathcal{G}=(\mathbf{0}~~\mathbbm{1}_9)$ is a $9\times10$ matrix, and
\begin{equation}
\mathcal{H}_9=\left(
                \begin{array}{ccc}
                  -e\mathbf{p}\cdot \mathbf{E}^\textsf{T}/\omega^2(\mathbf{p}) & -2\mathbf{p}\times & -2m \\
                  -2\mathbf{p}\times & \mathbf{0} & \mathbf{0} \\
                  2(m^2+\mathbf{p}\cdot \mathbf{p}^\textsf{T})/m & \mathbf{0} & \mathbf{0} \\
                \end{array}
              \right)\bigg|_{\mathbf{p}\rightarrow \mathbf{q}-e\mathbf{A}(t)}.\nonumber
\end{equation}
Thus, we can get the one-particle momentum distribution function $f(\mathbf{q},t)$ by solving Eq. (\ref{eq7}) with the initial conditions $f(\mathbf{q},-\infty)=\mathbbm{w}_9(\mathbf{q},-\infty)=0$. Integrating the distribution function over full momenta at $t\rightarrow+\infty$, we have the number density of created pairs
\begin{equation}\label{eq8}
n(+\infty)=\int\frac{d^3q}{(2\pi)^3}f(\mathbf{q},+\infty).
\end{equation}

\begin{figure}[htbp]\suppressfloats
\includegraphics[width=15cm]{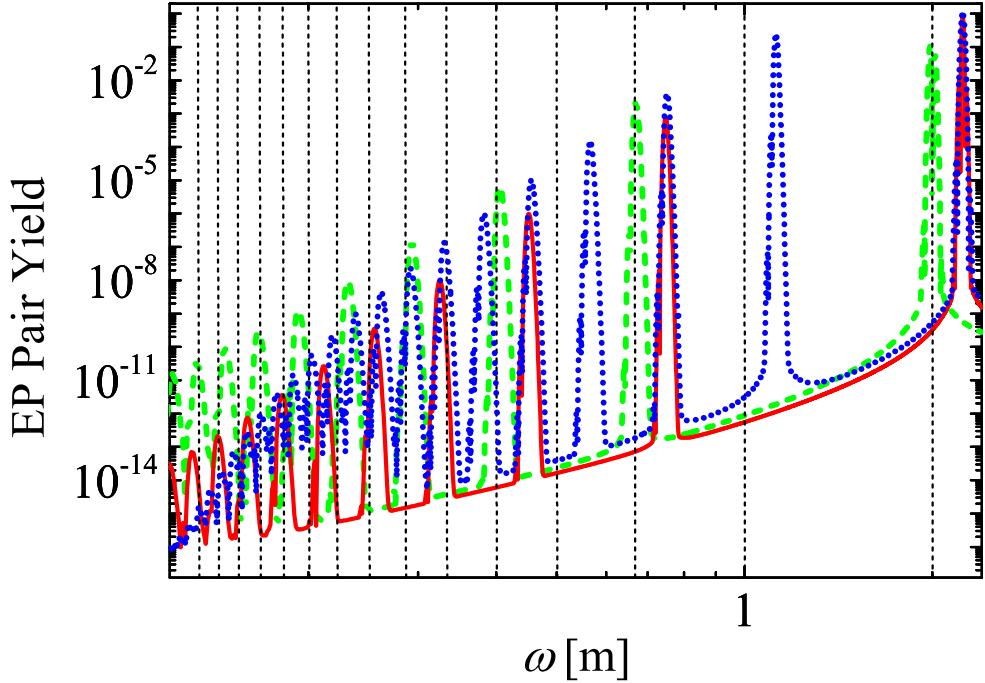}
\caption{\label{Fig1}(color online) Log-log plot of the EP pair yield as a function of the frequency $\omega$. The dashed green line for $q_x=q_y=q_z=\delta=0$, the solid red line for $q_x=q_z=\delta=0,q_y=0.5$, and the dotted blue line for $q_y=q_z=0,q_x=0.5,\delta=1$. The vertical dashed lines denote the peak positions simply estimated by the equation $n\omega=2m$ for $n$-photon thresholds. The electric field parameters are chosen as $E_0=0.1$, $\tau=100$.}
\end{figure}

\textit{Effective energy.}---Here we introduce an effective energy to interpret the mass shift effects appeared in the nonperturbative multiphoton pair production [cf. Fig. \ref{Fig1}]. Based on the Dirac sea picture, there is an energy gap $2m$ between the negative- and the positive-energy states. Therefore, the laser frequency needed for $n$-photon pair production can be simply estimated via the energy conservation equation $n\omega=2m$. However, the simplified estimate will become very rough when the external fields are considered, because the original energy gap can be deformed by the fields.  In addition, it is well known that in intense laser-matter interactions, the object's energy landscape can be modified by the ponderomotive energy which is defined as the cycle averaged oscillation energy of the electron in an oscillating electric field \cite{Quesnel}. Inspired by these, we propose an effective energy, the root-mean-square of the electron's total energy
\begin{equation}\label{eq9}
\Omega_{\mathrm{rms}}=\sqrt{\Big\langle \Big(\sqrt{m^2+[\mathbf{q}-e\mathbf{A}(t)]^2}\Big)^2\Big\rangle}
\end{equation}
with the average over a laser cycle $\langle\rangle$, to modify the original energy conservation equation and achieve $n\omega=2\Omega_{\mathrm{rms}}$ which determines the laser frequency needed for $n$-photon pair production. We emphasize that our effective energy can also be seen as the effective mass from the viewpoint of replacing the original energy gap described by the rest mass of electrons. More specifically, for the external field (\ref{eq1}), the effective energy becomes
\begin{equation}\label{eq10}
\Omega_{\mathrm{rms}}=m\sqrt{1+q^2+\frac{1+\delta^2}{2}\frac{e^2 E_0^2}{m^2\omega^2}},
\end{equation}
with $q=(q_x^2+q_y^2+q_z^2)^{1/2}/m$. Note that we do not consider the effect of the pulse shape because it is unimportant for a long pulse duration in the multiphoton process. From Eq. (\ref{eq10}), we can see that the energy gap depend not only on the momenta of created pairs but also on the polarizations of electric fields. This is verified by the DHW solutions in Fig. \ref{Fig1}. Further, we can get the shift laser frequency
\begin{equation}\label{eq11}
\omega_n=\sqrt{\frac{2(1+q^2)m^2}{n^2}+\sqrt{\frac{4(1+q^2)^2 m^4}{n^4}
+\frac{2(1+\delta^2)e^2E_0^2}{n^2}}}
\end{equation}
needed for $n$-photon pair production by equation $\omega_n=2\Omega_{\mathrm{rms}}/n$, and then the shift mass $M$ by equation $M=n\omega_n/2$.

\begin{figure}[htbp]\suppressfloats
\includegraphics[width=15cm]{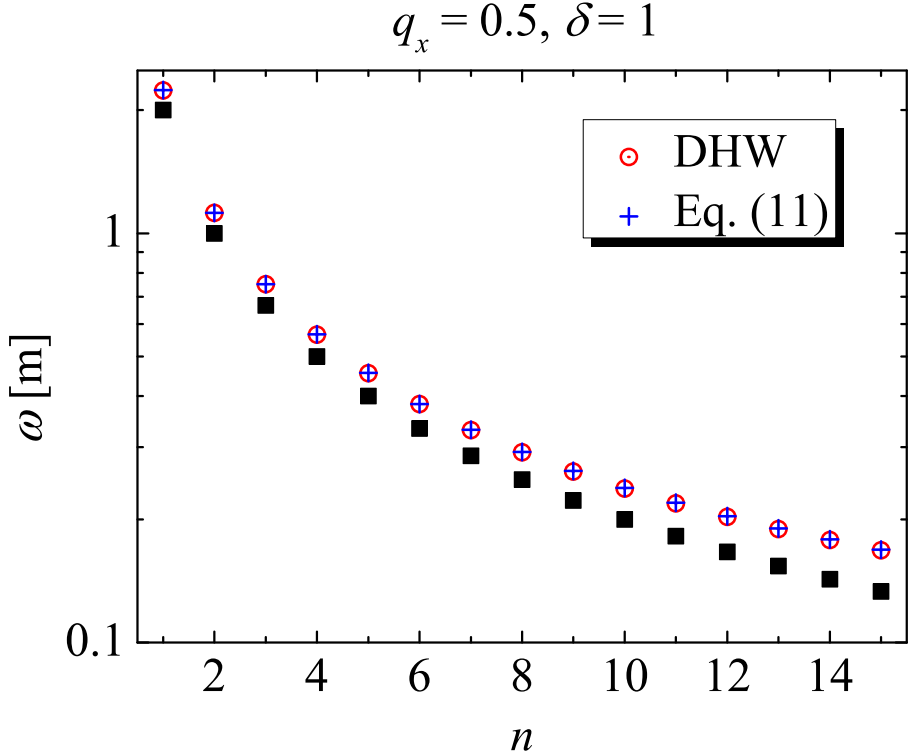}
\caption{\label{Fig2}(color online) The frequency $\omega$ as a function of the photon number $n$ for the peak of the EP pair yield in $n$-photon process. The black squares are the results simply estimated by the equation $n\omega=2m$. The red cycles are the solutions from DHW formalism, and the blue plus signs are the theoretical predictions of mass shift effects from Eq. (\ref{eq11}). Here $q_x=0.5$ and $\delta=1$. The electric field parameters are the same as in Fig. \ref{Fig1}.}
\end{figure}

\begin{figure}[htbp]\suppressfloats
\includegraphics[width=15cm]{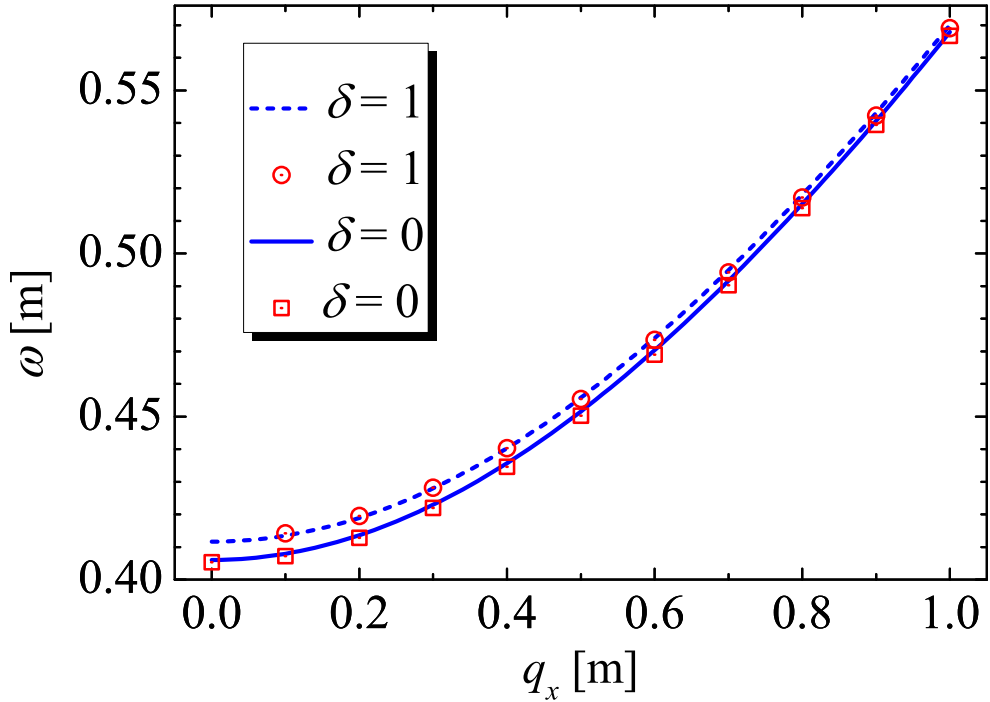}
\caption{\label{Fig3}(color online) The frequency $\omega$ as a function of the longitudinal momentum $q_x$ for the five-photon peak of the EP pair yield. The blue lines are the theoretical predictions of mass shift effects from Eq. (\ref{eq11}), and the red symbols are the solutions from DHW formalism. The upper line and symbol are the results for $\delta=1$. The lower ones are the results for $\delta=0$. The electric field parameters are the same as in Fig. \ref{Fig1}.}
\end{figure}

\begin{figure}[htbp]\suppressfloats
\includegraphics[width=15cm]{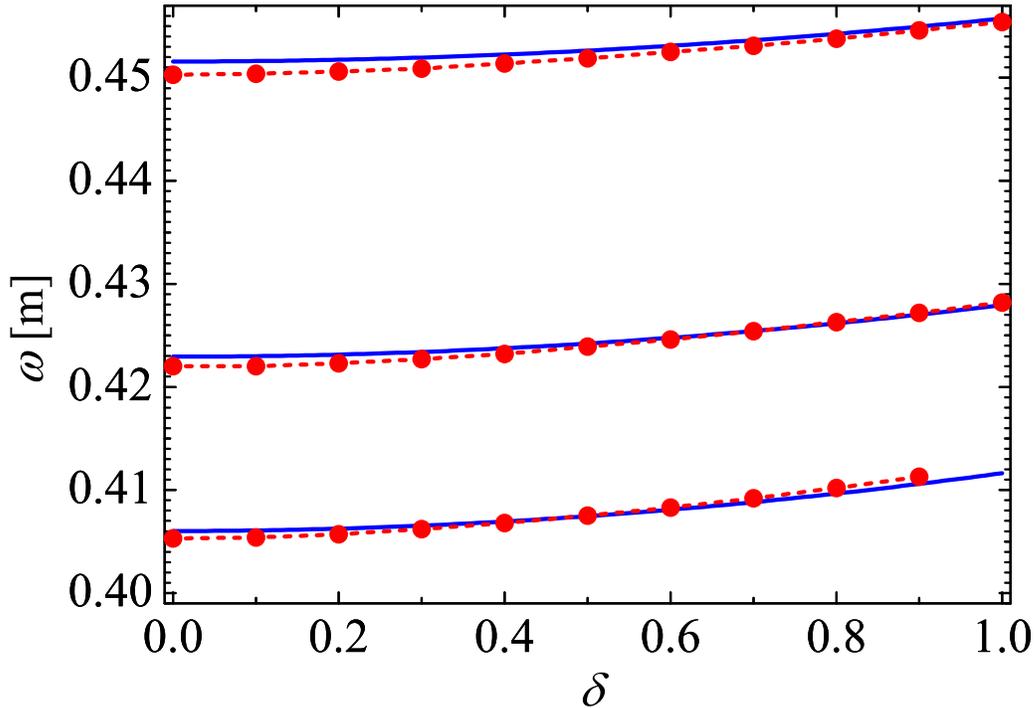}
\caption{\label{Fig4}(color online) The frequency $\omega$ as a function of the electric field polarization $\delta$ for the five-photon peak of the EP pair yield. The solid blue lines are the theoretical predictions of mass shift effects from Eq. (\ref{eq11}), and the dashed red lines are the solutions from DHW formalism. The results from top to bottom are for $q_x=0.5$, $0.3$, and $0$, respectively. The electric field parameters are the same as in Fig. \ref{Fig1}.}
\end{figure}

\begin{figure}[htbp]\suppressfloats
\includegraphics[width=15cm]{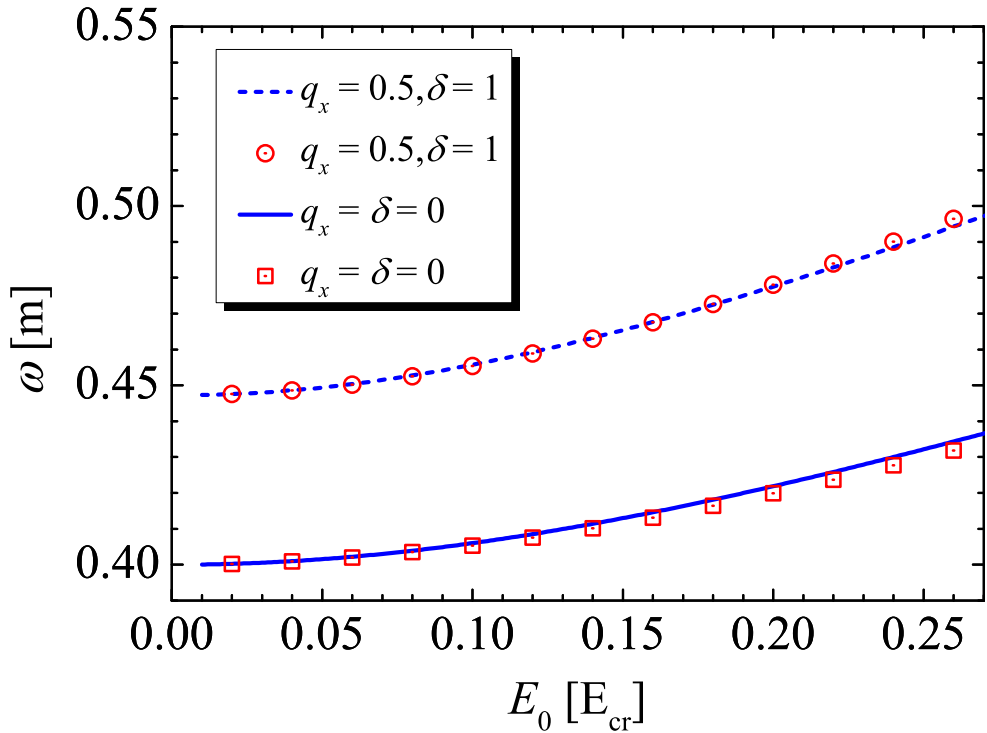}
\caption{\label{Fig5}(color online) The frequency $\omega$ as a function of the electric field strength $E_0$ for the five-photon peak of the EP pair yield. The blue lines are the theoretical predictions of mass shift effects from Eq. (\ref{eq11}), and the red symbols are the solutions from DHW formalism. The upper line and symbol are the results for $q_x=0.5$ and $\delta=1$. The lower ones are for $q_x=\delta=0$. The electric field parameters are chosen as $\tau=100$.}
\end{figure}

\begin{figure}[htbp]\suppressfloats
\includegraphics[width=15cm]{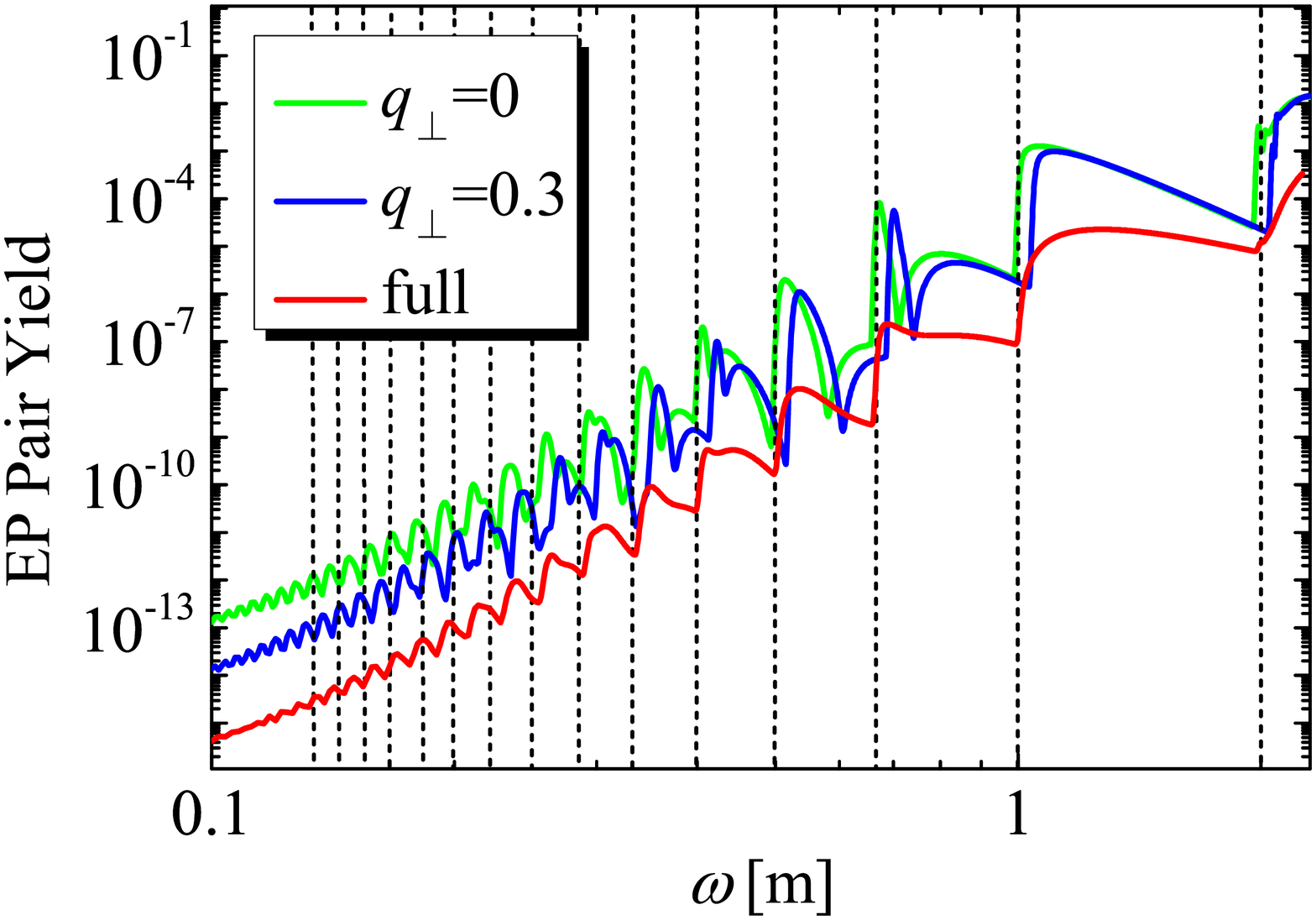}
\caption{\label{Fig6}(color online) Log-log plot of the EP pair yield as a function of the frequency $\omega$, for the transverse momentum $q_\perp=(q_y^2+q_z^2)^{1/2}/m=0$ (upper green line), $0.3$ (middle blue line), and the full momentum space (lower red line). The vertical dashed black lines denote the peak positions simply estimated by the equation $n\omega=2m$ for $n$-photon thresholds. Here $\delta=0$ and the electric field parameters are the same as in Fig. \ref{Fig1}.}
\end{figure}

\textit{Results and discussions.}---In Fig. \ref{Fig1}, we have found that the mass shift effects depend on both the momenta of created EP pairs and the polarizations of electric fields. Furthermore, we can find that the dashed green line shows the similar results as in Refs. \cite{Ruf,Mocken}, namely, there are no peaks of the EP pair yield at even photon number for vanishing particle momenta. And these have been interpreted as the parity-selection rule based on the conservation of charge-conjugation parity. Here we emphasize that the condition of zero momenta is indispensable for a general polarization electric field. For a linear polarization electric field, $\delta=0$, however, the results still hold as long as the longitudinal momenta vanish (see the solid red line). For a circular polarization electric field, $\delta=\pm1$, the vanishing momentum condition leads to zero resonance of the EP pair yield. This novel result can be explained from the definition of the total energy of electrons $\Omega(\mathbf{q},t)$, because the condition of zero momenta makes the energy gap described by $2\Omega(\mathbf{0},t)$ independent of the laser frequency.

Now let us calculate the corresponding laser frequencies $\omega(n,q,\delta,E_0)$ for the peak values of EP pair yield via the estimate of Eq. (\ref{eq11}) and compare them with the DHW solutions. For convenience, we fix $q_y=q_z=0$.  Figure \ref{Fig2} shows the laser frequency changing with the photon number $n$ for the peak of EP pair yield with $q_x=0.5$ and $\delta=1$. It is found that the momenta of created pairs indeed affect the threshold estimate, i.e., as the momenta increase the higher laser frequency is need for the $n$-photon process. We can also see that the estimates of Eq. (\ref{eq11}) are agreement with the numerical solutions of DHW formalism (\ref{eq7}) very well. In Fig. \ref{Fig3}, we show the relation between the laser frequency needed for $5$-photon pair production (corresponding to $\gamma=4$) and the momenta of created EP pairs for $\delta=0$ and $\delta=1$. It is shown that the laser frequency offsets rapidly grow with the increasing particle momenta. Moreover, the circular polarization electric fields have a larger shift laser frequency than the linear ones and the difference between them is reduced by increasing the momenta of created EP pairs. The effects of the polarizations of electric fields on the threshold estimate of $5$-photon pair production for $q_x=0.5$, $0.3$, and $0$ (from top to bottom) are depicted in Fig. \ref{Fig4}. It shows that the results of our estimate (\ref{eq11}) are consistent with the numerical solutions of DHW formalism (\ref{eq7}). Additionally, one can find that the consistency of the theoretical predictions and the numerical solutions has small changes for different particle momenta. To fully verify the threshold estimate of Eq. (\ref{eq11}), we also study the effects of electric field strength on the laser frequency needed for $5$-photon pair production. The results for $q_x=\delta=0$ and $q_x=0.5,\delta=1$ are shown in Fig. \ref{Fig5}. It can be found that the estimate of Eq. (\ref{eq11}) (red symbols) fit the solutions of DHW formalism (blue lines) well, especially for a low field strength. More importantly, from Figs. \ref{Fig4} and \ref{Fig5}, we find that the role of the polarizations of electric fields played in determining the peak positions of $n$-photon process is different from that of the electric field strength, though it seems that there is no difference between them from Eq. (\ref{eq11}). This difference can also be seen from the discussion about Figure \ref{Fig1} aforementioned and more details will be reported elsewhere.

Figure \ref{Fig6} shows the total EP pair yield calculated from Eq. (\ref{eq8}) as a function of the laser frequency for the fixed transverse momenta $q_\perp=0$, $0.3$,  and the full momentum space with a linear polarization electric field $\delta=0$. One can see that the solutions of DHW formalism (green line) can give the same result as in Fig. 1 of Ref. \cite{Kohlfurst2014}. For a fixed transverse momentum, the integration of the distribution function over the longitudinal momenta gives a complex structure of the EP pair yield changing with the laser frequency. However, the corresponding laser frequency of the peak values of $n$-photon pair production can still be estimated from Eq. (\ref{eq11}) by replacing $q$ with the transverse momentum $q_\perp$, i.e., the longitudinal momentum is simply set to zero since the number of created EP pairs near zero longitudinal momenta dominate the final results, especially for a small laser frequency. Unfortunately, when we further integrate the distribution function over the transverse momenta, it is found that the $n$-photon thresholds of frequency cannot be precisely estimated by Eq. (\ref{eq11}) with either zero transverse momentum or a fixed value of $q$. Obviously an important reason of the disagreement of the exact numerical results to the theoretical estimation of Eq.(\ref{eq11}) for the full momenta integrated number density should be the coupling between the longitudinal and the transverse momenta, which is worthy to investigate further in the future.

By the way the extensive studies we have made (not presented here) shows that the estimates of $n$-photon thresholds from Eq. (\ref{eq11}) are sill hold true for the EP pair yield by integrating $q_x$ with a fixed $q_\perp$ for the other polarization electric fields $\delta\neq0$. Certainly for the total EP pair yield by integrating the full momentum space, the difficulty of theoretical estimation of Eq. (\ref{eq11}) is also exist in case of $\delta\neq0$ as in $\delta=0$. This means the results obtained above, on the one hand, indicate that the mass shift effects can be greatly changed by integrating over the full momentum space so that the theoretical analysis becomes more complex. However, on the other hand, they manifest that the mass shift effects can be clearly presented in the momentum spectra, even if the EP yield is for the situation where the longitudinal momentum is integrated but the transverse momentum is fixed.

In a summary, we have investigated the mass shift effects in nonperturbative multiphoton pair production for arbitrary polarized electric fields both theoretically by proposing an effective energy and numerically by using the real-time DHW formalism. It is found that the theoretical results are well consistent with the numerical ones.  Moreover, the important roles of the momenta of created EP pairs and the polarizations of external electric fields played in the mass shift effects are investigated for the first time. These results are valuable to deepen the understanding of nonperturbative multiphoton mechanism. The pronounced mass shift effects are useful to distinguish the EP pair production from other background processes and can be detected in the experiments underway. A full theoretical analysis for the number density by integrating distribution function through full momentum space is still an open problem.

\textit{Acknowledgements.}---This work was supported by the National Natural Science Foundation of China (NSFC) under Grant No. 11475026, No. 11175023 and No.11335013, and also supported partially by the Open Fund of National Laboratory of Science and Technology on Computational Physics at IAPCM and the Fundamental Research Funds for the Central Universities (FRFCU). The computation was carried out at the HSCC of the Beijing Normal University.

\end{document}